\documentclass[reprint,showpacs,showkeys,preprintnumbers,amsmath,amssymb,aps,prl]{revtex4-1}
\usepackage{graphicx}
\usepackage{float}
\usepackage{color}

\begin{document}

\preprint{Ouardi et al., Spin Gapless Semiconductors}

\title{Realization of spin gapless semiconductors: \\
       the Heusler compound Mn$_2$CoAl.}

\author{Siham~Ouardi}
\affiliation{Max Planck Institute for Chemical Physics of Solids,
             D-01187 Dresden, Germany}
             
\author{Gerhard~H.~Fecher}
\email{fecher@cpfs.mpg.de}
\affiliation{Max Planck Institute for Chemical Physics of Solids,
             D-01187 Dresden, Germany}

\author{J{\"u}rgen~K{\"u}bler}
\affiliation{Institut f{\"u}r Festk{\"o}rperphysik, Technische Universit{\"a}t Darmstadt, 
             D-64289 Darmstadt, Germany}

\author{Claudia~Felser}
\affiliation{Max Planck Institute for Chemical Physics of Solids,
             D-01187 Dresden, Germany}

\date{\today}

\begin{abstract}

Recent studies have reported an interesting class of semiconductor materials 
that bridge the gap between semiconductors and halfmetallic ferromagnets. These 
materials, called {\it spin gapless semiconductors}, exhibit a bandgap in one 
of the spin channels and a zero bandgap in the other and thus allow for tunable spin transport.
 Here, a theoretical and experimental study of the spin gapless Heusler compound Mn$_2$CoAl is presented. 
It turns out that Mn$_2$CoAl is a very peculiar {\it ferrimagnetic semiconductor} 
with a magnetic moment of 2~$\mu_B$ and a high Curie temperature of 720~K.
Below 300~K, the compound exhibits nearly temperature-independent 
conductivity, very low, temperature-independent carrier concentration, and
a vanishing Seebeck coefficient. The magnetoresistance changes sign with 
temperature. In high fields, it is positive and non-saturating at low 
temperatures, but negative and saturating at high temperatures. The anomalous Hall 
effect is comparatively low, which is explained by the close antisymmetry of the 
Berry curvature for $k_z$ of opposite sign.

\end{abstract}



\maketitle

%%%%%%%%%%%%%%%%%%%%%%%%%%%%%%%%%%%%%%%%%%%%%%%%%%%%%%%%%%%%%%%%%%%%%%
%\section{Introduction}

\begin{figure}[htb]
\begin{center}
\includegraphics[width=6cm]{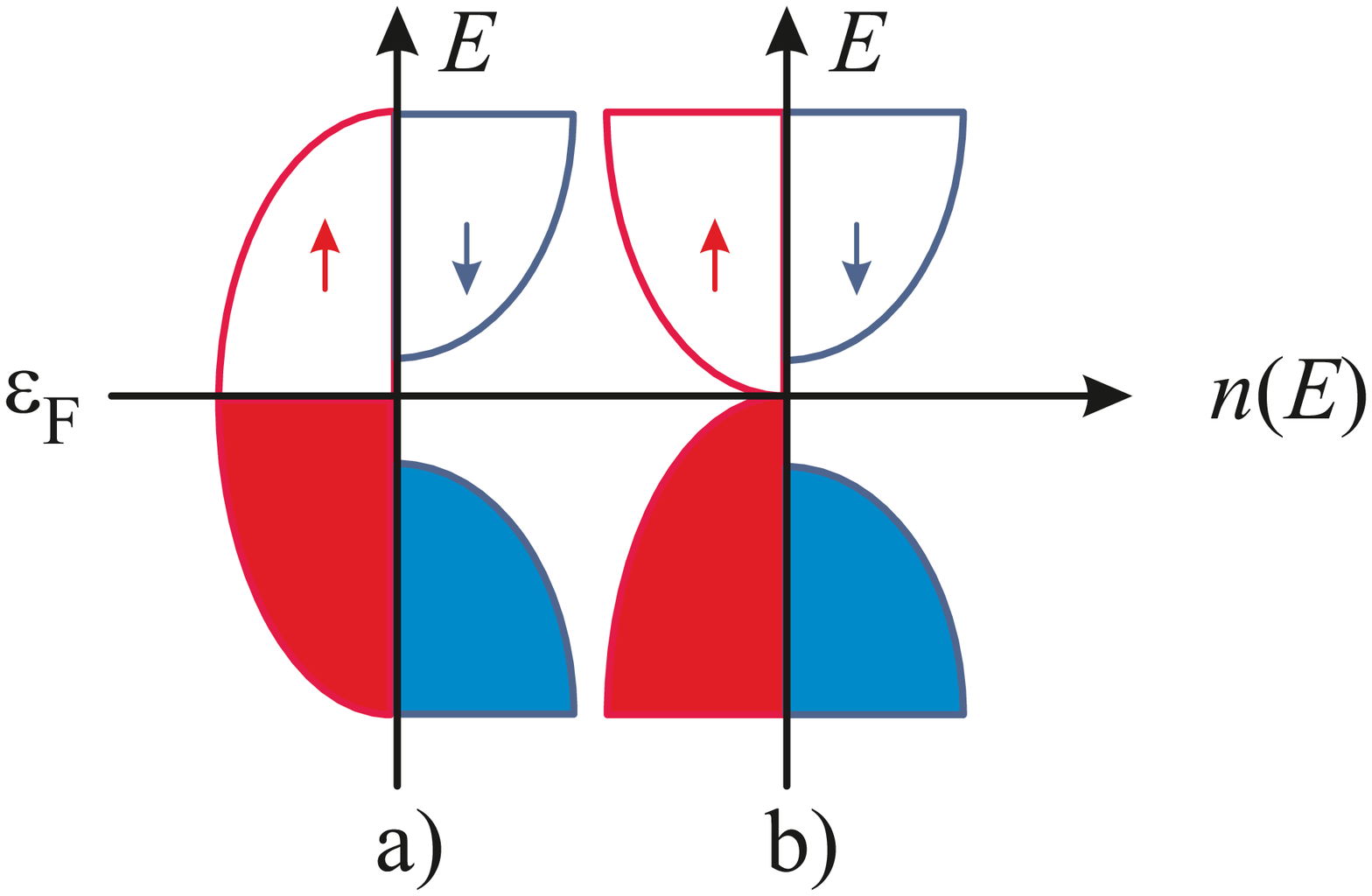}
\caption{Density of states schemes. \\
         The schematic density of states $n(E)$ as a function of energy $E$ is shown for
         a) a halfmetallic ferromagnet and 
         b) a spin gapless semiconductor. The occupied states are indicated by filled areas.
         Arrows indicate the majority ($\uparrow$) and minority ($\downarrow$) states.}
\label{fig:scheme}
\end{center}
\end{figure}

Recent studies have reported an interesting class of materials, namely {\it 
spin gapless} semiconductors~\cite{Wan08}. These materials are closely 
related to halfmetallic ferromagnets~\cite{GME83}, the properties of which also 
bridge the gap between metals and semiconductors. Special cases of  
halfmetallic materials are halfmetallic ferrimagnets with antiparallel 
orientation of their magnetic moments. This results in some cases in completely 
compensated halfmetals~\cite{WKF06c}, which are sometimes termed halfmetallic 
antiferromagnets~\cite{LGr95}. In all halfmetallic materials, transport is 
mediated by electrons having only one kind of spin, i.e., the minority or  
majority spin. {\it Spin gapless} semiconductors exhibit an additional 
phenomenon, namely an open bandgap in one spin channel and a closed bandgap 
({\it zero bandgap}) in the other. The principal density of states schemes for 
halfmetallic ferromagnets and spin gapless semiconductors are shown in 
Figure~\ref{fig:scheme}.

%%%%%%%%%%%%%%%%%%%%%%%%%%%%%%%%%%%%%%%%%%%%%%%%%%%%%%%%%%%%%%%%%%%%%%
%%%%%%%%%%%%%%%%%%%%%%%%%%%%%%%%%%%%%%%%%%%%%%%%%%%%%%%%%%%%%%%%%%%%%%

In the present example, the Fermi energy intersects the majority density in 
halfmetallic ferromagnets (Figure~\ref{fig:scheme}a)). The class of spin gapless 
semiconductors appears if the conduction and valence band edges of the majority 
electrons touch at the Fermi energy $\epsilon_F$ (Figure~\ref{fig:scheme}b)). 
This class is thus an important sub-class of {\it zero bandgap} insulators or 
{\it gapless semiconductors}~\cite{Tsi96}. In zero bandgap materials, no 
threshold energy is required to move electrons from occupied states to empty 
states. These materials therefore exhibit unique properties, as their 
electronic structures are extremely sensitive to external influences~\cite{Tsi96}. 
Many of the known gapless semiconductors exhibit a so-called inverted band 
structure~\cite{GBP67,Tsi96}, and they act as topological 
insulators~\cite{MNZ04,KMe05}, in which the bandgap is re-opened in the presence 
of a strong spin--orbit (SO) interaction. In {\it spin gapless} semiconductors, not 
only the excited electrons but also the holes can be 100\% spin polarized. 
Indeed, the Fermi energy needs to fall inside the majority gap, and it cannot 
touch the band edges. This results in unique transport properties that can be 
influenced by the magnetic state of the material or by magnetic fields.

%%%%%%%%%%%%%%%%%%%%%%%%%%%%%%%%%%%%%%%%%%%%%%%%%%%%%%%%%%%%%%%%%%%%%%

Based on these findings of previous studies, as mentioned above, the Heusler 
compound Mn$_2$CoAl was investigated in greater detail, using theoretical and 
experimental methods, in order to prove the occurrence of spin gapless 
semiconductivity. The compound crystallizes in the inverse Heusler structure 
with space group $F\:\overline{4}3m$ and Wyckhoff sequence $acbd$ for the atoms 
Mn-Mn-Co-Al. To explain the electronic structure and magnetic properties, {\it 
ab initio} calculations were performed using the augmented spherical wave (ASW) 
method~\cite{WKG79}, which is extremely fast and efficient. Berry curvatures were 
computed  using ASW density functional calculations and the 
wave functions that they provide, allowing the anomalous Hall conductivity to be
obtained from the Berry curvatures. The technical details of such calculations 
have been reported in Reference~\cite{KF12}.

%%%%%%%%%%%%%%%%%%%%%%%%%%%%%%%%%%%%%%%%%%%%%%%%%%%%%%%%%%%%%%%%%%%%%%
\begin{figure}[htb]
   \begin{center}
   \includegraphics[width=9cm]{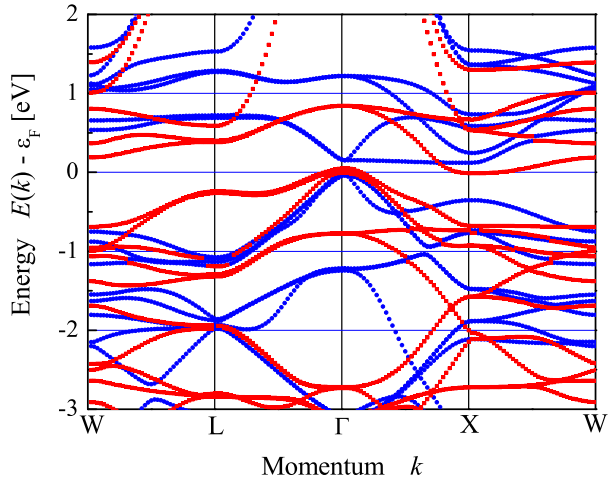}
   \caption{Band structure of Mn$_2$CoAl. \\
            The band structure calculated with spin--orbit interaction is shown.
            Majority (red) and minority (blue) spin characters of the bands are distinguished
            by different colors.}
   \label{fig:bands}
   \end{center}
\end {figure}
%%%%%%%%%%%%%%%%%%%%%%%%%%%%%%%%%%%%%%%%%%%%%%%%%%%%%%%%%%%%%%%%%%%%%%

Figure~\ref{fig:bands} shows the band structure of Mn$_2$CoAl. The calculations 
were performed with SO interactions respected for all atoms, as this 
is essential for the Berry curvature calculations reported below. The electronic 
structure agrees well with reports of bare spin density functional 
calculations not respecting SO~\cite{LDL08}. In Figure~\ref{fig:bands}, the spin 
character of the bands is indicated by different colors. The spin character of 
the bands close to the Fermi energy is pure, even though the SO 
interaction mixes, in general, states with different spin projections. 
Obviously, the band structure also exhibits the expected halfmetallic character with 
a bandgap in the minority states (blue) when the SO interaction is included in 
the calculations.  A most interesting observation is the bandgap that also 
appears in the majority channel. This is neither a semiconductor nor a 
halfmetallic ferromagnet, because the indirect gap appears to be just closed as a result of 
bands touching the Fermi energy at $\Gamma$ (valence band) and $X$ 
(conduction band). The {\it zero bandgap} behavior of the minority electrons 
converts the {\it halfmetallic ferrimagnet} into a {\it spin gapless 
semiconductor}. The magnetic moment is calculated to be 2~$\mu_B$, as observed 
experimentally (see below). The site-resolved spin magnetic moments are 
approximately $m_{\rm Mn(4a)}=-2\mu_B$, $m_{\rm Co(4b)}=+1\mu_B$, and $m_{\rm 
Mn(4c)}=+3\mu_B$, confirming a ferrimagnetic character with antiparallel 
coupling of the moments at the neighboring Mn sites. In the following, it will be 
shown that various extraordinary physical properties are in agreement with the 
proposed behavior of a {\it spin gapless semiconductor}. 

\begin{figure}[htb]
\centering
\includegraphics[width=9cm]{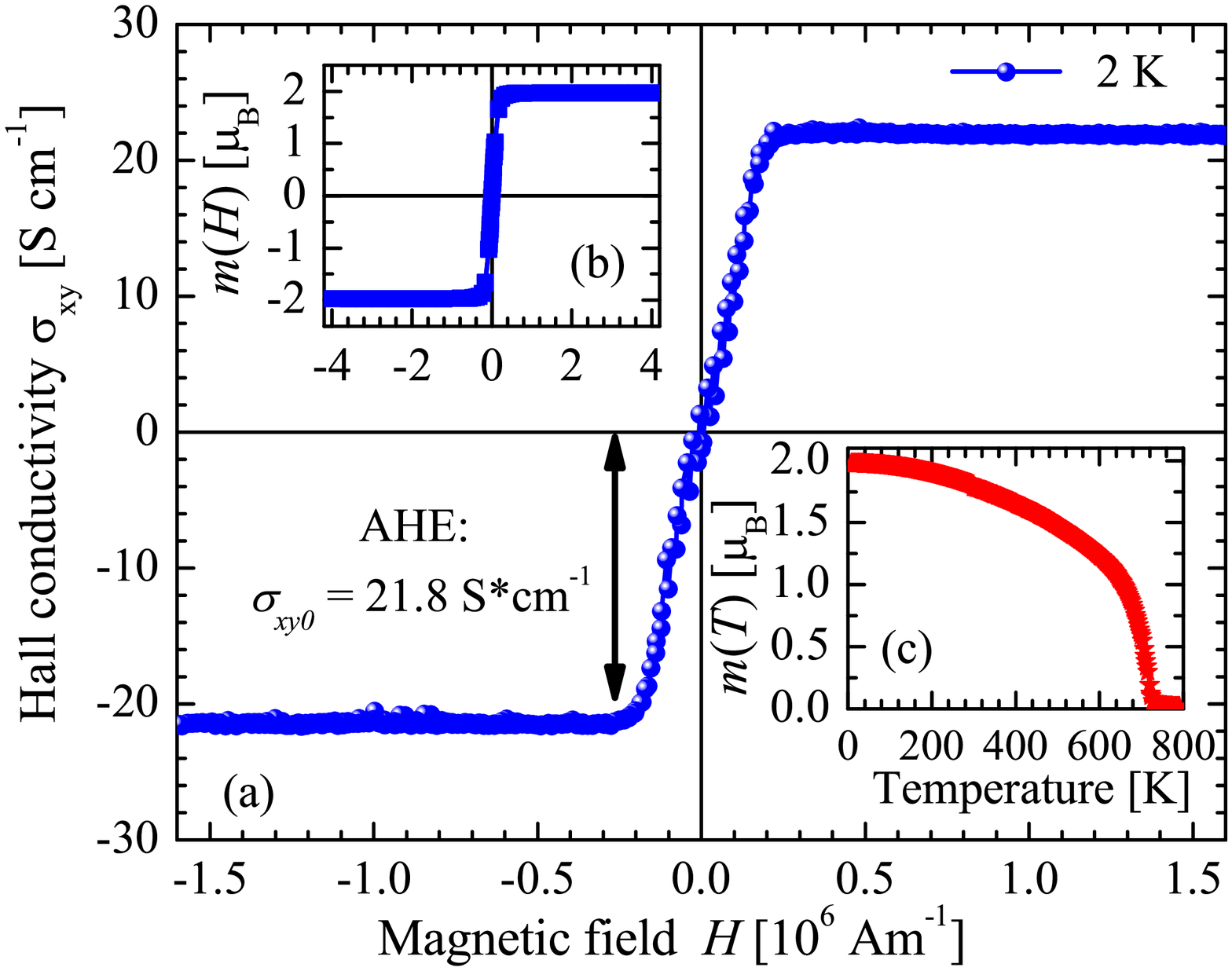}
\caption{Magnetic properties and anomalous Hall effect of Mn$_2$CoAl. \\
        The Hall conductivity $\sigma_{xy}$ is shown as a function of the applied
        magnetic field. The anomalous Hall effect is defined by the difference between $\sigma_{xy}$
        values in zero and saturation fields ($\approx 0.3$~T).
        Inset (a) shows the hysteresis $m(H)$ measured at 2~K.
        Inset (b) shows the temperature dependence of the magnetization $m(T)$ measured in a field of 1~T.}
\label{fig:AHE}
\end{figure}

Polycrystalline Mn$_2$CoAl was prepared by arc melting. The resulting ingots 
were annealed at different temperatures under argon in quartz ampules. The 
annealing was followed by fast cooling to 273~K (quenching in ice-water). The 
crystalline structures, homogeneities, and compositions of the samples were checked 
by X-ray diffraction (XRD) and energy-dispersive X-ray (EDX) spectroscopy. No 
impurities or inhomogeneities were detected by EDX. Transport measurements were 
performed using a physical properties measurement system (PPMS; Quantum 
Design). For transport measurements, samples of $(2 \times 2 \times 8)$~mm$^3$ 
were cut from the ingots. The temperature was varied from 1.8~K to 350~K. The 
Hall effect of the compound was measured in the temperature range from 5~K to 
300~K in magnetic induction fields from -9 to +9~T. The Hall coefficient was 
calculated from the slope of the measured Hall coefficient $R_H$. The carrier 
concentration $n=1/e R_H$ was extracted from $R_H$ using a single-band 
model~\cite{Ro06} ($e$ is the elementary electronic charge). The temperature 
dependence of the magnetization was measured from 4.8 K to 800~K using a 
magnetic properties measurement system (MPMS; Quantum Design). For these 
measurements, small sample pieces of approximately 8~mg were used. The 
experiments on Mn$_2$CoAl confirmed that it crystallizes in the expected 
crystalline structure. Powder XRD revealed single-phase samples with the Li$_2$AgSb 
structure and a lattice parameter of $a=5.798$~{\AA} for the post-annealed 
samples. The saturation magnetic moment is 2~$\mu_B$ at low temperature 
($T<5$~K) (see inset b in Figure~\ref{fig:AHE}) and the compound has a Curie 
temperature of 720~K (Figure~\ref{fig:AHE}c).

%%%%%%%%%%%%%%%%%%%%%%%%%%%%%%%%%%%%%%%%%%%%%%%%%%%%%%%%%%%%%%%%%%%%%%%%%%%%%%%
%\subsection{experimental results}

\begin{figure}[htb]
\includegraphics[width=7cm]{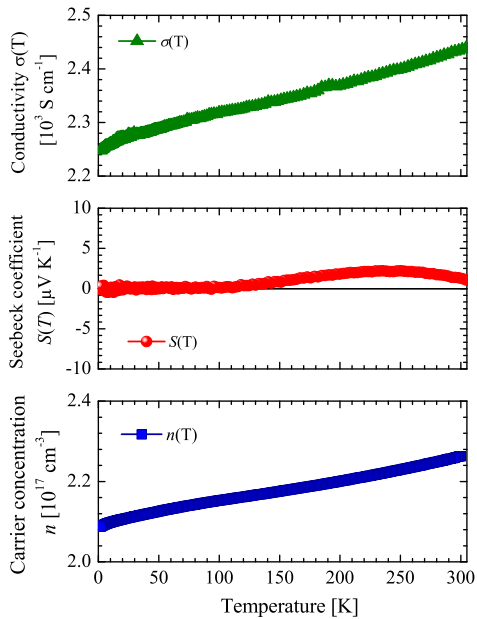}
\caption{Temperature dependences of conductivity $\sigma_{xx}(T)$, 
         Seebeck coefficient $S(T)$, and carrier concentration $n(T)$ of Mn$_2$CoAl.}
\label{fig:transp}
\end{figure}

The temperature dependences of the electrical conductivity $\sigma(T)$, Seebeck 
coefficient $S(T)$, and carrier concentration $n(T)$ of Mn$_2$CoAl are shown in 
Figure~\ref{fig:transp}. The conductivity $\sigma(T)$ clearly exhibits 
non-metallic behavior, and is of semiconducting type: it increases with 
increasing temperature and a value of about 2440~S/cm is obtained at 300~K. As 
shown in Figure~\ref{fig:transp}(b), the Seebeck coefficient $S(T)$ nearly 
vanishes in the temperature range from 5 K to 150 K and adopts a very low value of 
only about 2~$\mu$V/K at 300~K. Measurement of the Hall coefficient shows
a very low carrier concentration of only $2\times10^{17}$~cm$^{-3}$ at 
300~K. It is also nearly constant with temperature and approaches a value as low 
as $1.3\times10^{17}$~cm$^{-3}$ at 2~K.

%%%%%%%%%%%%%%%%%%%%%%%%%%%%%%%%%%%%%%%%%%%%%%%%%%%%%%%%%%%%%
%%%%%%%%%%%%%%%%%%%%%%%%%%%%%%%%%%%%%%%%%%%%%%%%%%%%%%%%%%%%%

The linear temperature coefficient of the resistivity is $-1.4\times10^{-9}$~$\Omega\cdot$m/K 
and nearly independent of temperature up to 300~K. This 
behavior is remarkable and not comparable to those of regular metals or semiconductors, 
which exhibit exponential increases or decreases in conductivity. As Mn$_2$CoAl 
is a well-ordered compound without antisite disorder, this behavior cannot be 
attributed to impurity scattering. A similar, linear behavior of the resistance 
was also reported for halfmetallic ferromagnets, but only at low 
temperatures~\cite{MM94}. The resistivity in those metallic systems 
($<0.5$~$\mu\Omega\cdot$m for PtMnSn) is, however, three orders of magnitude 
lower than that of Mn$_2$CoAl ($\approx400$~$\mu\Omega\cdot$m). The 
temperature-independent charge-carrier concentration is typical for gapless 
systems~\cite{Tsi96}. The value observed here is of the same order as that observed 
for HgCdTe ($10^{15}-10^{17}$~cm$^{-3}$) and considerably lower than that for 
Fe$_2$VAl ($10^{21}$~cm$^{-3}$)~\cite{OAG07}, a proposed semiconducting Heusler 
compound. Also, for gapless systems, the Seebeck effect is expected---as 
observed here---to vanish over a wide range of temperatures as a result of  
electron and hole compensation.

%%%%%%%%%%%%%%%%%%%%%%%%%%%%%%%%%%%%%%%%%%%%%%%%%%%%%%%%%%%%%
\begin{figure}[htb]
\centering
\includegraphics[width=6cm]{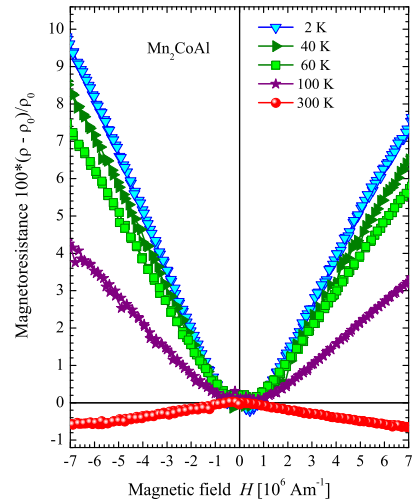}
\caption{Magnetoresistance of Mn$_2$CoAl measured at different temperatures.}
\label{fig:MR}
\end{figure}
%%%%%%%%%%%%%%%%%%%%%%%%%%%%%%%%%%%%%%%%%%%%%%%%%%%%%%%%%%%%%

So far, the electronic transport properties, together with the predicted and 
measured magnetic moment of 2~$\mu_B$, support the view that Mn$_2$CoAl is a 
spin gapless semiconductor. To investigate further transport properties of the 
spin gapless state, the magnetoresistance (MR) and anomalous Hall conductivity 
of Mn$_2$CoAl were measured.

The results of the MR measurements at different temperatures are displayed in 
Figure~\ref{fig:MR}, and show a remarkable effect. Above 200~K, the MR is low 
and exhibits a negative, saturating dependence on the applied magnetic field. At 
lower temperatures, the MR becomes positive and reaches a value of about 10\% at 
2~K for an induction field of 9~T. The low-temperature MR is clearly 
non-saturating and nearly linear, even in high fields. Above ~1 T, its derivative has 
an approximately constant value of $10^{-4}$~T$^{-1}$ at 2 K. A change of sign 
appears at around 150~K, where the Seebeck effect vanishes, and remains the same at lower 
temperatures. Various gapless semiconductors have already been shown to exhibit 
quantum linear MR~\cite{Abr98} in very small transverse magnetic 
fields. For the present compound, and for spin gapless materials in general, the 
behavior in low fields becomes more complicated as a result of the influence of the 
ferrimagnetic order and the non-saturated magnetization in low fields. The MR 
changes sign at low fields, as seen from Figure~\ref{fig:MR}; an appropriate theory is needed
to describe the MR in spin gapless systems. 

%%%%%%%%%%%%%%%%%%%%%%%%%%%%%%%%%%%%%%%%%%%%%%%%%%%%%%%%%%%%%
%%%%%%%%%%%%%%%%%%%%%%%%%%%%%%%%%%%%%%%%%%%%%%%%%%%%%%%%%%%%%

The anomalous Hall conductivity $\sigma_{xy}=\frac{\rho_{xy}}{\rho^2_{xx}}$ was 
extracted from the magnetic-field-dependent transport measurements to 
disentangle the low-field behavior in more detail. The result is shown in 
Figure~\ref{fig:AHE}. $\sigma_{xy}(H)$ follows the field dependence of the 
magnetization $m(H)$ (inset b of Figure~\ref{fig:AHE}). The anomalous Hall 
conductivity $\sigma_{xy0}$ has a very low value of 22~S/cm at 2~K. This value 
is about 20 to 100 times smaller than those of other halfmetallic Heusler 
compounds~\cite{BBV12}.

\begin{figure}[htb]
   \begin{center}
   \includegraphics[width=9cm]{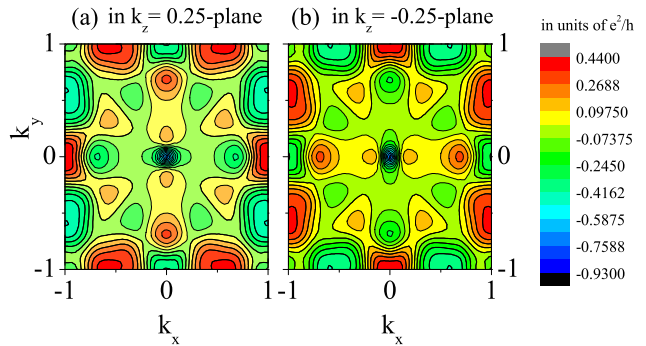}
   \caption{Berry curvature in the $k_z = 0.25$ plane 
            (a) and $k_z = -0.25$ plane (b) for Mn$_{2}$CoAl.}
   \label{fig:berry}
   \end{center}
\end {figure}

The anomalous Hall effect was calculated using the Berry-phase 
approach~\cite{KF12} in order to explain its extraordinarily small value. The 
result of the calculation is shown in Figure~\ref{fig:berry}. The symmetry of 
the Berry curvature $\Omega_z(\mathbf k)$ in a plane of the Brillouin zone is 
clearly visible. This pattern symmetry is different from the patterns of regular 
ferromagnets (e.g., compare Fe~\cite{YKM04}). Moreover, it is easily seen that 
the pattern for momentum vectors with the opposite sign (here at $k_z=\pm0.25$) is 
nearly antisymmetric. The anomalous Hall conductivity is given by a Brillouin 
zone integral over the Berry curvature $\Omega_z(\mathbf k)$ over all occupied 
states\cite{YKM04}:

\begin{equation}\label{eq1}
  \sigma_{xy}=\frac{e^2}{h}
  \,\int\frac{d\mathbf{k}}{(2\pi)^d}
  \Omega_z(\mathbf{k})f(\mathbf{k}),
\end{equation}

where $f(\mathbf k)$ is the Fermi distribution function (at $T=0$) and the 
dimension $d=3$. This makes clear that the nearly vanishing anomalous Hall 
effect of Mn$_2$CoAl arises from the {\it antisymmetry} of the Berry curvature 
for $k_z$ vectors of opposite sign. Indeed, a numerical integration of equation 
(\ref{eq1}) gives $\sigma_{xy}= 3$~S/cm, which is lower than the experimental 
value. This small value comes from positive and negative contributions of 
about 150~S/cm each. The small switching field used in the 
experiment could therefore account for the difference between the calculated and experimental values.

%%%%%%%%%%%%%%%%%%%%%%%%%%%%%%%%%%%%%%%%%%%%%%%%%%%%%%%%%%%%%%%%%%%%%%
%%%%%%%%%%%%%%%%%%%%%%%%%%%%%%%%%%%%%%%%%%%%%%%%%%%%%%%%%%%%%%%%%%%%%%

%%%%%%%%%%%%%%%%%%%%%%%%%%%%%%%%%%%%%%%%%%%%%%%%%%%%%%%%%%%%%%%%%%%%%%
%\section{Summary}

Ab initio calculations suggested that the Heusler compound Mn$_2$CoAl is a good 
candidate for spin gapless semiconductivity. This unique class of materials will 
have a considerable impact on the field of spintronics as it opens up new and 
advanced possibilities for physical phenomena and devices based on spin 
transport. In spin gapless materials, only one spin channel contributes to the 
transport properties, whereas the other spin channel allows for tunable 
charge-carrier concentrations. In the present work, Mn$_2$CoAl was synthesized. Its 
crystalline structure is of the inverse Heusler type. The saturation magnetic 
moment is 2~$\mu_B$ at 5~K and the Curie temperature of 720~K makes it suitable 
for applications. It has been shown by experiments and calculations that 
Mn$_2$CoAl is a spin gapless semiconductor with nearly temperature-independent, 
low conductivity of the order of $2\times10^{5}$~S/m, and a low charge-carrier 
concentration of the order of $10^{17}$~cm$^{-3}$, as well as a vanishing 
Seebeck coefficient. The temperature dependence of the MR is 
non-trivial. In high fields, it is positive, nearly linear, and non-saturating, with 
a value of 10\% at 2~K in a magnetic field of 9~T. At temperatures above 150~K, 
it is negative and saturating, with low values. Berry curvature calculations were 
used to explain the low value of the anomalous Hall effect of 22~S/cm at 2~K.

%%%%%%%%%%%%%%%%%%%%%%%%%%%%%%%%%%%%%%%%%%%%%%%%%%%%%%%%%%%%%%%%%%%%%%
\bigskip
\begin{acknowledgments}

The authors thank W. Schnelle (MPI, Dresden) for performing part of the 
transport measurements. This work was financially supported by the {\it Deutsche 
Forschungs Gemeinschaft} DfG (projects TP~1.2-A and TP~2.3-A of Research Unit 
FOR~1464 {\it ASPIMATT}).

\end{acknowledgments}

%%%%%%%%%%%%%%%%%%%%%%%%%%%%%%%%%%%%%%%%%%%%%%%%%%%%%%%%%%%%%%%%%%%%%%
%\bibliography{ouardi_spingapless}


\end{document}